\documentclass[aps,amsmath,twocolumn,amssymb,titlepage,10pt]{revtex4-1}
\usepackage[T1]{fontenc}
\usepackage[utf8]{inputenc}
\usepackage{amsmath}
\usepackage{braket}
\usepackage{amsfonts}
\usepackage{comment}
\usepackage{graphicx}
\usepackage{tikz}
\usepackage{hyperref}

\usepackage{amssymb}
\usepackage{amsmath} 
\usepackage{mathtools}
\DeclareMathOperator{\Tr}{Tr}
\newcommand{\newbrace}[1][]{
\begin{tikzpicture}[baseline=-0.5ex]
\draw[#1] (0,0) -- (0.3,0.3);
\draw[#1] (0,0) -- (0.3,-0.3);
\end{tikzpicture}
}
\newenvironment{casesnew}[1][->]%
{\;\newbrace[#1]\;\begin{array}{@{}l@{}}}%
{\end{array}}
\newcommand{\underbrack}{\underbracket[0.5pt][1.5pt]}

\begin{document}
\title{Hierarchical single-ion anisotropies in spin-1 Heisenberg\\ antiferromagnets on the honeycomb lattice}
\begin{abstract}
We examine the thermal properties of the spin-1 Heisenberg antiferromagnet on the honeycomb lattice
in the presence of an easy-plane single-ion anisotropy as well as the effects of an  additional weak in-plane  easy-axis anisotropy. In particular, using large-scale quantum Monte Carlo simulations, we analyze the scaling of the correlation length near the thermal phase transition into the ordered phase. This allows us to quantify the temperature regime above the critical point in which -- in spite of the additional in-plane easy-axis anisotropy -- characteristic easy-plane physics, such as near a Berezinskii-Kosterlitz-Thouless transition, can still be accessed. 
Our theoretical analysis is motivated by recent neutron scattering studies of the spin-1 compound BaNi${}_2$V${}_2$O${}_8$ in particular, and it addresses basic quantum spin models for  generic spin-1 systems with  weak anisotropies, which we probe over the full range of experimentally relevant correlation length scales. 
\end{abstract}
\author{Nils Caci}
\author{Lukas Weber}
\author{Stefan Wessel}
\affiliation{Institute for Theoretical Solid State Physics, RWTH Aachen University, JARA Fundamentals of Future Information Technology, and \\ JARA Center for Simulation and Data Science, 52056 Aachen, Germany}
\maketitle

\section{Introduction}\label{Sec:Intro}
In recent years, the search for solid-state realizations of Berezinskii-Kosterlitz-Thouless (BKT) 
topological phase transitions~\cite{Berezinskii1971, Kosterlitz1973, Kosterlitz1974} in magnetic compounds  has
lead to the identification of several quasi two-dimensional (2D) antiferromagnetic candidate materials~\cite{Opherden2020,Tutsch2014,Hu2020,Shen2019, Li2020,LiX2020}; for an overview over the earlier literature  on  magnetic compounds for which BKT transitions have been considered, cf. Ref.~\cite{Taroni2008}.
While in most systems, the BKT behavior is obstructed by the presence of residual interlayer couplings, these were found to be negligible for the specific Ni${}^{2+}$ based  compound BaNi${}_2$V${}_2$O${}_8$, in which  spin-1 degrees of freedom reside in effectively decoupled 2D honeycomb lattice layers~\cite{Rogado2002,Klyushina2017}. 
Instead of interlayer coupling, a weak easy-plane single-ion anisotropy stabilizes dominant planar (XY)  correlations in BaNi${}_2$V${}_2$O${}_8$ upon lowering the temperature below about 80~K \cite{Klyushina2017}. It was found in theoretical studies that  weakly anisotropic 2D Heisenberg antiferromagnets indeed exhibit a vortex-driven BKT transition at a temperature $T_\mathrm{BKT}$ set by the Heisenberg exchange coupling, separating a disordered high-temperature regime from a quasi long-range ordered phase below $T_\mathrm{BKT}$~\cite{Cuccoli2003}.

However, BaNi${}_2$V${}_2$O${}_8$ features a true antiferromagnetic ordering transition~\cite{Rogado2002}  at a N\'eel temperature $T_\mathrm{N}$ of about 47.75~K~\cite{Klyushina2021}.
Detailed inelastic neutron scattering studies of the  low-temperature ordered state of BaNi${}_2$V${}_2$O${}_8$ furthermore indicate the presence of an additional, though very weak, anisotropy which favors the alignment of the magnetic moments along only a subset of directions within the spin's easy-plane~\cite{Klyushina2017}.

As a most basic model system for the magnetism of BaNi${}_2$V${}_2$O${}_8$, which accounts for these essential properties~\cite{Klyushina2017,Klyushina2021}, we consider here the Hamiltonian
\begin{equation}\label{eq:ham}
H=J\sum_{\langle i,j \rangle} \mathbf{S}_i \cdot \mathbf{S}_j + D_z \sum_i (S^z_i)^2 - D_x\sum_i (S^x_i)^2,
\end{equation}
in terms of spin-1 degrees of freedom $\mathbf{S}_i$ residing on the sites of a honeycomb lattice with an antiferromagnetic  nearest-neighbor exchange constant $J>0$ (i.e., the first sum extends over all nearest-neighbor bonds). 
Further (weak) interaction terms, e.g.,  between next-nearest neighboring spins were considered in Ref.~\cite{Klyushina2017}
based on a linear spin wave theory modeling.  The more basic  model $H$ in Eq.~(\ref{eq:ham}) was then later found to also  account well for the neutron scattering data on BaNi${}_2$V${}_2$O${}_8$, with the estimated values of 
$J=8.8$ meV, 
$D_z=0.099$ meV and $D_x=0.0014$ meV~\cite{Klyushina2021}.

The weak anisotropy $D_z\ll J$, along with an even weaker $D_x < D_z$, indeed entails a hierarchy of single-ion anisotropies:  A finite $D_z>0$ leads to the preferred orientation of the spin moments within the spin-XY plane at low temperatures, while the additional $D_x>0$ favors their alignment in the spin-X direction. Correspondingly, in the pure easy-plane limit, $D_x=0$, the Hamiltonian $H$ has a residual O(2) symmetry in the spin-XY plane and exhibits a BKT transition at a finite transition temperature $T_\mathrm{BKT}$ 
(as quantified in detail below). 
On the other hand, a finite value of $D_x>0$ explicitly breaks the spin symmetry of $H$ down to a discrete Z${}_2$ symmetry in the spin-X direction, and in this case the system  instead exhibits a 2D Ising ordering transition at a finite N\'eel temperature $T_\mathrm{N}$ (also quantified below).

Based on the underlying lattice structure, the in-plane anisotopy in BaNi${}_2$V${}_2$O${}_8$ may be argued to  exhibit a 6-fold symmetry instead of a single in-plane easy-axis direction~\cite{Klyushina2021}. Due to the irrelevancy of a Z${}_6$ perturbation at the BKT transition~\cite{Jose1977}, the BKT transition would then not be affected by the weak in-plane anisotropy 
(in the opposite limit of the classical Z${}_6$-symmetric clock model the transition was instead found to  no longer be of  BKT type~\cite{Lapilli06}). However, the microscopic models that were derived from the inelastic neutron scattering data contain an explicit Z${}_2$ symmetric in-plane anisotropy, as in Eq.~(\ref{eq:ham}), which is a strongly relevant perturbation at the BKT transition. Moreover, it was observed in Ref.~\cite{Klyushina2021} that quantum effects in BaNi${}_2$V${}_2$O${}_8$ need to be accounted for  in order to quantitatively model the correlations in this compound, even though the underlying thermal physics is dominated by classical fluctuations in the weak anisotropy regime.
Therefore, our study
focuses on  the question, whether for temperatures close to and  above $T_\mathrm{N}$, one may still be able to identify in the magnetic correlations characteristic features of the dominant easy-plane anisotropy $D_z$, i.e., remnants of the BKT physics that govern the magnetism 
of the quantum spin model in Eq.~(\ref{eq:ham}) in the  pure easy-plane limit  $D_x=0$.

In fact, it was observed recently that within a finite temperature window above $T_\mathrm{N}$, the magnetic correlation length $\xi$ in BaNi${}_2$V${}_2$O${}_8$  exhibits a temperature dependence that fits well to the BKT scaling formula $\xi(T)\propto \exp(\: b/\sqrt{T-T_\mathrm{BKT}}\: )$, \cite{Kosterlitz1974} as compared to conventional power-law scaling~\cite{Klyushina2021}. In the above, $b$ is a non-universal number, and $T_\mathrm{BKT}$ defines an (effective) BKT transition temperature, estimated for BaNi${}_2$V${}_2$O${}_8$ to be 44.7~K, i.e., $T_\mathrm{BKT}$ is below $T_\mathrm{N}$.
This indicates that  the BKT physics of vortex excitations still controls the initial buildup of the magnetic correlations in BaNi${}_2$V${}_2$O${}_8$ upon approaching the thermal phase transition, but the underlying BKT transition is  preempted by the N\'eel ordering transition that is induced by the additional weak in-plane anisotropy~\cite{Klyushina2021}.

Here, we assess the above scenario for the case of the effective model Hamiltonian $H$, which allows us to make  direct and quantitative comparisons of the correlation length scaling between the full hierarchical model and the pure easy-plane limit ($D_x=0$). For this purpose, we performed a series of large-scale quantum Monte Carlo (QMC) simulations of the hierarchical Hamiltonian $H$, using a variant of the stochastic series expansion (SSE) method~\cite{Sandvik1991,Sandvik1999,Syljuasen2002}. Having in mind a hierarchy of weak anisotropies that is appropriate for the  compound BaNi${}_2$V${}_2$O${}_8$, we  concentrate here on  the regime where $D_x< D_z\ll  J$. 
However, we found that important aspects for the analysis of BKT transitions in spin-1 systems on the honeycomb lattice are not available from  previous studies. For  this reason, we first consider in the following several limits of the  Hamiltonian $H$ and related models.

More specifically, in the following Sec.~\ref{Sec:XY}, we first examine the BKT transition of the spin-1 XY model on the honeycomb lattice, in order to set the stage for the later discussion of the hierarchical model $H$. 
Then, in Sec.~\ref{Sec:EP}, we concentrate on the pure easy-plane limit ($D_x=0$) of the Hamiltonian $H$, and then examine the full hierarchical model  
in Sec.~\ref{Sec:HA}. 
Final conclusions are then drawn in Sec.~\ref{Sec:Conclusions}.
Important technical aspects of the employed SSE algorithm that are specific to  the QMC simulation of the anisotropic Hamiltonian $H$ are provided in the appendix. There, we also examine in detail the quantum phase transition that emerges in the pure easy-plane model for  larger values of $D_z$. 
Finally, we  provide in the appendix  also an analysis of the pure easy-axis regime ($D_z=0$) of the Hamiltonian $H$, for which  we identify  an enhanced ordering temperature in the large-$D_x$ regime relative to the classical Blume-Capel limit.  

\section{The Spin-1 XY Model}\label{Sec:XY}
While BKT transitions in several anisotropic quantum spin systems have been studied to a high precision in the past, we are not aware of any detailed study of anisotropic spin-1 systems or even the most basic spin-1 XY model on the honeycomb lattice. Hence, before we examine the full  hierarchical Hamiltonian $H$, we first consider the identification of the BKT transition and the correlation length scaling in the most basic spin-1 honeycomb lattice model that exhibits a BKT transition, i.e., the spin-1 XY model. This model is  defined by the Hamiltonian
\begin{equation}
    H_{\mathrm{XY}} = J\sum_{\langle i,j \rangle} S_i^x S_j^x + S_i^y S_j^y , 
\end{equation}
which has a transverse antiferromagnetically ordered ground state for $J>0$ on a bipartite lattice, and a transverse ferromagnetic ground state for $J<0$. On a bipartite lattice, such as the honeycomb lattice, both cases can be related by a sublattice rotation,  so that here we need to  treat explicitly only one of these cases. We consider the antiferromagnetic case, in order to set up the notation in the following to apply also directly to the Hamiltonian $H$, which also contains an antiferromagnetic exchange interaction. 

In the following, we consider the honeycomb lattice in terms of a triangular lattice with a two-site unit cell. For the QMC simulations, we use finite $L \times L$ rhombi of $L^2$ unit cells and $N=2L^2$ spins, taking periodic boundary conditions in both lattice directions. Furthermore,  we denote the  lattice constant in terms of the distance between neighboring lattice sites by $a_0$.

\begin{figure}[t]
    \centering
    \includegraphics{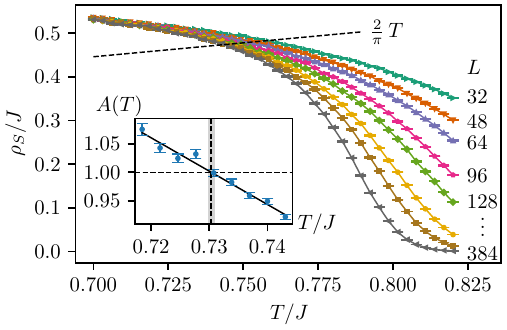}
    \caption{Spin stiffness $\rho_S$ for different system sizes $L$ as function of temperature $T$ of the spin-1 XY model on the honeycomb lattice. The dashed line denotes the scaling form of the universal jump. The inset shows the quantity $A(T)$ from the finite-size scaling analysis.  The critical point is denoted by the dashed vertical line, where $A(T)=1$ holds, obtained using a linear fit (solid line).}
    \label{fig:XY_rho}
\end{figure}

A standard means of identifying the BKT transition temperature $T_\mathrm{BKT}$ in O(2) symmetric  systems is based on the behavior of the spin stiffness $\rho_S$, which is  predicted to exhibit a universal jump of $\rho_S = 2\, T_\mathrm{BKT}/\pi$ at the system's BKT transition temperature $T_\mathrm{BKT}$~\cite{Nelson1977}. For the specific case of the XY model considered here, we will denote the BKT transition temperature by $T^\mathrm{XY}_\mathrm{BKT}$ in the following. 
Within the SSE QMC approach $\rho_S$ can be calculated from the spin winding number fluctuations \cite{Pollock1987,Sandvik1997}
\begin{equation}
    \rho_S = \frac{T}{2\,A_\mathrm{uc}} \bigl(\langle W_x^2\rangle + \langle W_y^2\rangle\bigr),
\end{equation}
where $W_x$ and $W_y$ are the total winding numbers in the orthogonal $x$ and $y$ direction, respectively. In order to compare to the universal scaling relation of the stiffness jump in the continuum limit, the winding number fluctuations are normalized by the unit cell area $A_\mathrm{uc}$ in units of $a_0^2$, which equals $A_\mathrm{uc}=\sqrt{3}/2$ for the  honeycomb lattice. 
To extract $T^\mathrm{XY}_\mathrm{BKT}$ from finite-size QMC data, we then follow the standard approach of Ref.~\cite{Harada1998}, which is based on the finite-size scaling form~\cite{Weber1988}
\begin{equation}
    \frac{\rho_S\,\pi}{2\,T} = A(T)\,\biggl( 1  + \frac{1}{2\,\log(L/L_0(T))}\biggr)
\end{equation}
that holds exactly at the transition point with $A(T_\mathrm{BKT})=1$. We  fitted this finite-size dependence to the data for different temperatures, using $A(T)$ and $L_0(T)$ as fit parameters. This allows us to accurately estimate the transition temperature, where $A(T_\mathrm{BKT})=1$ holds. Our results from this approach are shown in Fig.~\ref{fig:XY_rho}, and we obtain from this analysis an estimate of $T^\mathrm{XY}_\mathrm{BKT}/J=0.7303(4)$ for the spin-1 XY model on the honeycomb lattice.

\begin{figure}[t]
    \centering
    \includegraphics{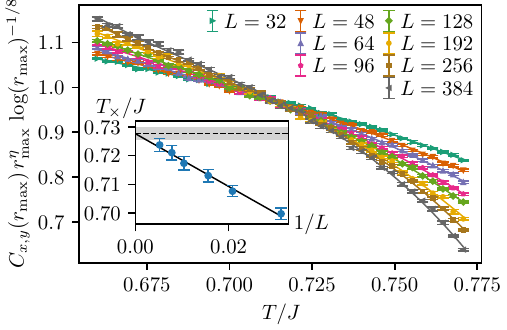}
    \caption{Transverse spin correlations $C_{x,y}(r_{\mathrm{max}})$, multiplied by ($r_\mathrm{max})^\eta\,\log(r_\mathrm{max})^{-1/8}$ for different system sizes $L$, as functions of $T$ of the spin-1 XY model on the honeycomb lattice. Near the crossing points they are approximated by polynomials of degree 3 (solid lines). The inset shows the temperature $T_\times$ of the crossing points between the fitting polynomials of linear system sizes $L$ and $2\,L$ as a function of inverse system size $1/L$, which is extrapolated to $1/L=0$ using a linear fit (solid line). }
    \label{fig:XY_CSS}
\end{figure}

As another approach to estimate $T^\mathrm{XY}_\mathrm{BKT}$, we analyze the transverse spin correlation function 
$C_{x}(r_{i,j})=\langle S_i^xS_j^x \rangle$, which for the XY model also equals $C_{y}(r_{i,j})=  \langle S_i^y S_j^y\rangle$, and where $r_{i,j}$ denotes the spatial distance between spins $i$ and $j$, accounting for the periodic boundary conditions. 
In the thermodynamic limit and at the BKT transition temperature, the magnitude $C_{x,y}(r)$ of these correlation functions is predicted to scale as
\begin{equation}
    C_{x,y}(r) \sim \frac{\ln(r)^{1/8}}{r^\eta}\biggl[1 + \mathcal{O}\biggl(\frac{\ln(r)^{1/8}}{r^\eta}\biggr)\biggr],
\end{equation}
with the critical exponent $\eta=1/4$~\cite{Amit1980}. We measured the values of $C_{x,y}(r_\mathrm{max}(L))$ at the largest available distance $r_\mathrm{max}(L)$ for different lattice sizes $L$. Based on the above scaling form, we can then estimate $T^\mathrm{XY}_\mathrm{BKT}$ from a crossing-point analysis of the appropriately rescaled values of $C_{x,y}(r_\mathrm{max}(L))$
between system sizes $L$ and $2\,L$, and performing an extrapolation to the thermodynamic limit ($1/L=0$), as shown in Fig.~\ref{fig:XY_CSS}.
Within the statistical uncertainty, the value  $T^\mathrm{XY}_\mathrm{BKT}/J=0.728(2)$ that we obtain for the BKT transition temperature from this analysis is in accord with the (more accurate) estimate based on $\rho_S$ reported above. 

Following the above identification of the BKT transition temperature in the XY-model, we next examine in detail  the behavior of the transverse spin correlation length $\xi^\mathrm{XY}$ in this model and its scaling behavior near $T^\mathrm{XY}_\mathrm{BKT}$. This  analysis will be important to our later study of the scaling behavior of the correlation length in the hierarchical model $H$.

To extract the correlation length from the spin correlations of a general spin model on the honeycomb lattice, we consider the  magnetic structure factor $S_\alpha(\mathbf{q})$, $\alpha=\{x,y,z\}$ defined as
\begin{equation}
    S_{\alpha}(\mathbf{q}) =\frac{1}{2}\left( S_\alpha^{AA}(\mathbf{q}) + S_\alpha^{BB}(\mathbf{q}) - S_\alpha^{AB}(\mathbf{q}) - S_\alpha^{BA}(\mathbf{q})\right),
\end{equation}
capturing the antiferromagnetic alignment inside the unit cell at $\mathbf{q}=\mathbf{0}=(0,0)$, 
where 
\begin{equation}
S_\alpha^{\mu \nu}(\mathbf{q}) = \frac{1}{L^2}\sum \limits_{m,n} e^{i\mathbf{q}\cdot(\mathbf{r}_m-\mathbf{r}_n)}C_\alpha^{\mu \nu }(\mathbf{r}_m-\mathbf{r}_n),
\end{equation}
with $m,n$ summed over the $L^2$ unit cells, 
is given in terms of  the  correlation function $C_\alpha^{\mu \nu}(\mathbf{r})$ between the $\alpha$ component of two spins at lattice sites belonging to sublattices $\mu,\nu\in\{A,B\}$, and where $\mathbf{r}$ denotes the separation of the unit cells with respect to the underlying triangular lattice.
The spin correlation length $\xi_\alpha$ of the fluctuations in the $\alpha$ direction is then obtained in the standard way \cite{Sandvik2010} as
\begin{equation}
    \xi_\alpha = \frac{1}{\sqrt{15/16}} \frac{1}{|\mathbf{q}_1|} \sqrt{\frac{S_\alpha(\mathbf{0})}{S_\alpha(\mathbf{q}_1)}-1},
\end{equation}
where $\mathbf{q}_1$ is one of the reciprocal lattice vectors closest to $\mathbf{0}=(0,0)$ on the $L\times L$ lattice, and the factor $1/\sqrt{15/16}$ is introduced to relate the estimator to the Ornstein-Zernike correlation length \cite{Ornstein1914}.

\begin{figure}[t]
    \centering
    \includegraphics{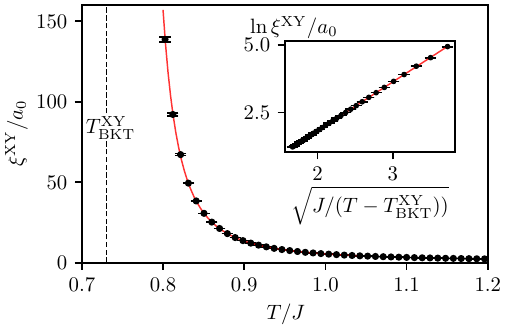}
    \caption{Correlation length $\xi^\mathrm{XY}$ of the spin-1 XY model on the honeycomb lattice as a function of $T$.
    The solid red line shows a fit to the BKT scaling formula. 
    The inset shows the same data on a logarithmic scale as a function of $1/\sqrt{T - T^\mathrm{XY}_\mathrm{BKT}}$.}
    \label{fig:XY_xis}
\end{figure}

\begin{figure}[t]
    \centering
    \includegraphics{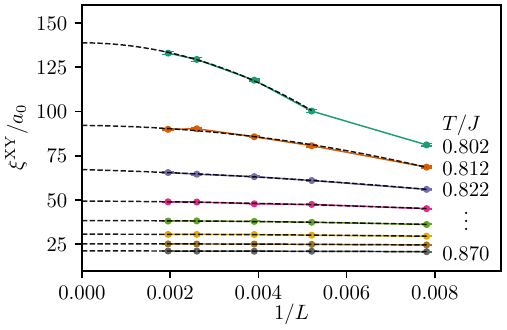}
    \caption{Extrapolation of the correlation length $\xi^\mathrm{XY}$ of the spin-1 XY model on the honeycomb lattice from the finite-size data for the lower temperatures from Fig.~\ref{fig:XY_xis}, as obtained from  fitting to polynomials of order 2 (dashed lines).}
    \label{fig:XY_xi_extrap}
\end{figure}

For the spin-1 XY model, we consider the correlation length of the transverse fluctuations, $\xi_x=\xi_y$, which we denote by $\xi^\mathrm{XY}$ in the following.
Its temperature dependence is shown in Fig.~\ref{fig:XY_xis}, as obtained from extrapolating the finite-size estimates to the thermodynamic limit (cf. Fig.~\ref{fig:XY_xi_extrap}). In this way, we are able to reliably extract values of $\xi^\mathrm{XY}$ up to about 140 lattice constants $a_0$. 

Close to the BKT temperature, the correlation length is predicted to scale as 
\begin{equation}\label{eq:BKTxiscaling}
\xi^\mathrm{XY} = a \exp \left(b/\sqrt{T - T^\mathrm{XY}_\mathrm{BKT}}\right),
\end{equation}
where  $a$  and $b$ are non-universal parameters. We find that the numerical data fits well to this BKT scaling form, as shown by the fit line in Fig.~\ref{fig:XY_xis}. In particular, we can also obtain from this analysis a further estimate of $T^\mathrm{XY}_\mathrm{BKT}/J =0.7305(3)$, consistent with our previous values.  
A more direct comparison to the BKT scaling form is obtained by examining $\xi^\mathrm{XY}$ on a logarithmic scale, for which the BKT scaling form yields 
\begin{equation}\label{eq:BKTxiscalinglog}
    \ln (\xi^\mathrm{XY}/a_0) = \ln(a/a_0)+  b/\sqrt{T - T^\mathrm{XY}_\mathrm{BKT}},
\end{equation}
which indeed fits well to the numerical data, as seen in the inset of Fig.~\ref{fig:XY_xis}. We thus find that for the spin-1 XY model on the honeycomb lattice, the correlation length closely follows the BKT scaling form upon approaching the BKT transition temperature.

\section{Pure Easy-Plane Regime}\label{Sec:EP}
After having examined the basic spin-1 XY model on the honeycomb lattice, we next turn our attention to the  easy-plane limit ($D_x=0$) of the Hamiltonian $H$, i.e., we consider
\begin{equation}
    H_\mathrm{EP}=J\sum_{\langle i,j \rangle} \mathbf{S}_i \cdot \mathbf{S}_j + D_z \sum_i (S^z_i)^2 .
\end{equation}
Here, a finite value of $D_z>0$ breaks the O(3) symmetry of the Heisenberg model down to a residual O(2) symmetry in the spin-XY plane, and in this two-dimensional model this is expected to 
lead to a BKT transition in the easy-plane, even for very weak anisotropies.
As already stated in Sec.~\ref{Sec:Intro},
here we focus on the regime of weak $D_z\ll J$, and we consider in detail  the value of $D_z=0.01J$,
which is of the order of   the value estimated for  BaNi${}_2$V${}_2$O${}_8$~\cite{Klyushina2017,Klyushina2021}.
In this regime, the model indeed exhibits an XY ordered antiferromagnetic ground state and a BKT transition into the low-T critical regime. 
In the following, we first identify the emergence of the BKT transition for the easy-plane spin-1 model on the honeycomb lattice and then analyze the correlation length scaling upon approaching the BKT transition temperature. 

\begin{figure}[t]
    \centering
    \includegraphics{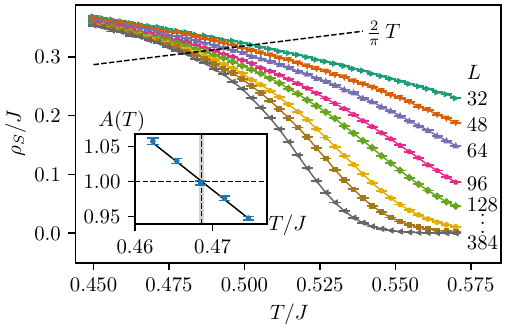}
    \caption{Spin stiffness $\rho_S$ for different system sizes $L$ as function of temperature $T$ for the spin-1 easy-plane model $H_\mathrm{EP}$  on the honeycomb lattice for $D_z=0.01J$. The dashed line denotes the scaling form of the universal jump. The inset shows the quantity $A(T)$ from the finite-size scaling analysis.  The critical point is denoted by the dashed vertical line, where $A(T)=1$ holds, obtained using a linear fit (solid line).}
    \label{fig:EP_rho}
\end{figure}

We  identity the BKT transition temperature using the spin stiffness $\rho_S$, following the same approach as introduced in Sec.~\ref{Sec:XY}. The result of this analysis for $H_\mathrm{EP}$ is shown in Fig.~\ref{fig:EP_rho}, from which we extract the value of $T^\mathrm{EP}_\mathrm{BKT}/J= 0.46860(1)$ (we verified that this estimate is also in accord with a corresponding analysis of the correlation function $C_{x,y}(r_\mathrm{max}(L))$, as for the spin-1 XY model in the previous section). 

\begin{figure}[t]
    \centering
    \includegraphics{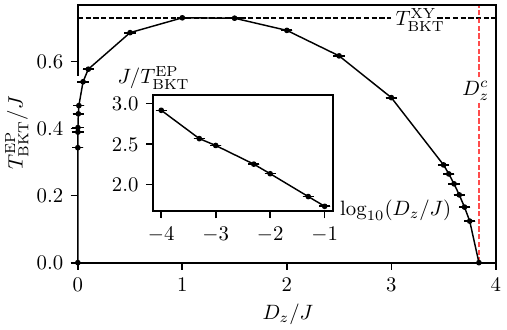}
    \caption{BKT transition temperature $T_\mathrm{BKT}^\mathrm{EP}$ of $H_\mathrm{EP}$ as a function of the easy-plane anisotropy $D_z$. The dashed vertical line denotes $D_z^c$ and the dashed  horizontal line the BKT transition temperature $T_\mathrm{BKT}^\mathrm{XY}$ of $H_\mathrm{XY}$. The inset indicates the logarithmic scaling of $1/T_\mathrm{BKT}^\mathrm{EP} \propto \ln(D_z/c)$ at small  $D_z$.}
    \label{fig:EP_phasediag}
\end{figure}

The above procedure can be repeated for varying values of $D_z$ in order to obtain a thermal phase diagram of the Hamiltonian $H_\mathrm{EP}$. For completeness, we present the QMC data for $T^\mathrm{EP}_\mathrm{BKT}$ as a function of $D_z$ in Fig.~\ref{fig:EP_phasediag}.
These results exhibit  several noticeable features: (i) as a function of   $D_z$, $T^\mathrm{EP}_\mathrm{BKT}$ exhibits a non-monotonous behavior: The initial increase of $T^\mathrm{EP}_\mathrm{BKT}$ with $D_z$ near the isotropic limit is followed by a reduction of $T^\mathrm{EP}_\mathrm{BKT}$ for $D_z \gtrsim 1$. (ii) $T^\mathrm{EP}_\mathrm{BKT}$ vanishes for $D_z$ approaching the value of $D_z^c\approx 3.8J$. In fact, as examined in more detail in App.~\ref{Sec:EP_app}, the Hamiltonian $H_\mathrm{EP}$ features a quantum phase transition within the three-dimensional (3D) XY universality class at $D_z^c/J= 3.83805(5)$, beyond which the XY-ordered antiferromagnetic ground state gets replaced by a non-magnetic state due to the proliferation of the local $S_i^z=0$ states for larger values of $D_z$. 
This quantum phase transition was also recently identified within a mean field approximation~\cite{Liu2020}, as well as by QMC for $H_\mathrm{EP}$ on a square lattice geometry~\cite{Zhang13}.
(iii) The maximum value of $T^\mathrm{EP}_\mathrm{BKT}$,  for $D_z\approx J$, is remarkably close to the value of the BKT transition in the spin-1 XY model $H_\mathrm{XY}$ (indicated by the horizontal line in Fig.~\ref{fig:EP_phasediag}). 
(iv) In the low-$D_z$ regime, we observe an approximate logarithmic suppression of $T^\mathrm{EP}_\mathrm{BKT}$, i.e.,
\begin{equation}
    T^\mathrm{EP}_\mathrm{BKT} \propto 1/\ln(D_z/c),
\end{equation}
where $c$ is a (non-universal) constant (cf. the inset of Fig.~\ref{fig:EP_phasediag}).
Such a leading logarithmic scaling was indeed obtained by  earlier 
spin-wave theory and renormalization group calculations~\cite{Hikami1980,Irkhin1999}, and was also observed in numerical studies  of both classical and spin-1/2  weakly anisotropic easy-plane XXZ models~\cite{Cuccoli2003}. 

For the easy-plane model $H_\mathrm{EP}$, the  correlation lengths $\xi_{x}$ and $\xi_y$  diverge upon approaching $T^\mathrm{EP}_\mathrm{BKT}$. Both quantify the transverse correlations and equal each other due to the residual O(2) symmetry. We thus denote this quantity by $\xi^\mathrm{EP}_{xy}$ in the following. We also consider the correlation length $\xi_z$ of the longitudinal fluctuations, which we denote by $\xi^\mathrm{EP}_{z}$  correspondingly. Both quantities were obtained as described in the previous section, based on the corresponding spin structure factors.

\begin{figure}[t]
    \centering
    \includegraphics{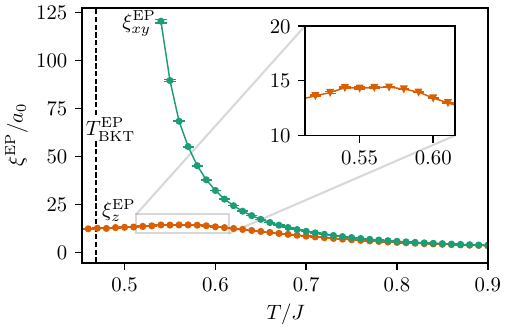}
    \caption{Correlation lengths $\xi^\mathrm{EP}_{xy}$ and $\xi^\mathrm{EP}_{z}$ as functions of temperature $T$ of the spin-1 easy-plane $H_\mathrm{EP}$ model on the honeycomb lattice for $D_z=0.01J$.}
    \label{fig:xiEP}
\end{figure}

The evolution of these correlation lengths with $T$, after an extrapolation to the thermodynamic limit,  is shown in Fig.~\ref{fig:xiEP}. Here, we again consider the value of $D_z=0.01J $. In addition to the expected increase of the transverse correlation length $\xi^\mathrm{EP}_{xy}$, we  observe a non-monotonous behavior in the longitudinal correlation length $\xi^\mathrm{EP}_{z}$: For temperatures larger than $T\approx 0.75 J$, both correlation lengths closely follow each other,
as expected from the O(3) symmetry of the leading Heisenberg exchange term in $H_\mathrm{EP}$. Below this scale however, 
$\xi^\mathrm{EP}_{xy}$ starts to deviate noticeably from $\xi^\mathrm{EP}_{z}$. Indeed, $\xi^\mathrm{EP}_{z}$, while initially still increasing upon lowering $T$, exhibits a broad maximum at a temperature $T_p$ of about $T_p\approx 0.56 J$ (see the inset in Fig.~\ref{fig:xiEP}), before it decreases slightly upon further lowering $T$. In earlier studies of spin-1/2 XXZ models on square lattice geometries,  similar behavior of the  correlation lengths was observed in the regime of  (weak) easy-plane exchange anisotropy~\cite{Cuccoli03}. There, the temperature of the maximum in the longitudinal correlation length was identified as a crossover scale separating the high-$T$ Heisenberg region from an intermediate temperature regime with enhanced  in-plane fluctuations above the BKT transition.

\begin{figure}[t]
    \centering
    \includegraphics{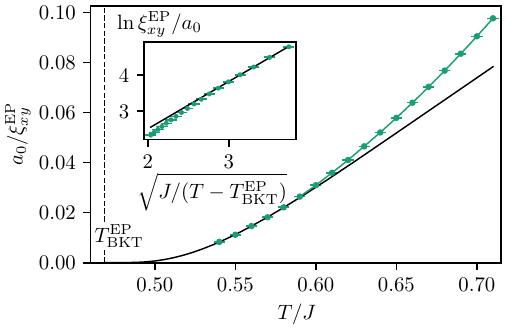}
    \caption{Inverse correlation length  $a_0/\xi^\mathrm{EP}_{xy}$ of the easy-plane model $H_\mathrm{EP}$ for  $D_z=0.01J$ as a function of  temperature in the vicinity of the BKT transition. The solid black  line is a fit of the exponential BKT scaling form to the lowest five data points. The inset shows the same data on a logarithmic scale as a function of $1/\sqrt{T - T^\mathrm{EP}_\mathrm{BKT}}$. }
    \label{fig:EPcoll}
\end{figure}

Returning to $H_\mathrm{EP}$, 
at about the crossover scale $T_p$
the further increase of the transverse correlation length $\xi^\mathrm{EP}_{xy}$ upon approaching  the 
BKT transition indeed starts to be well described in terms of the exponential BKT scaling  of Eq.~(\ref{eq:BKTxiscaling}), with $T^\mathrm{XY}_\mathrm{BKT}$ replaced by $T^\mathrm{EP}_\mathrm{BKT}$.
This is illustrated by the fits to the BKT scaling form in Fig.~\ref{fig:EPcoll}. The main panel of Fig.~\ref{fig:EPcoll} shows  the inverse transverse correlation length, $a_0/\xi^\mathrm{EP}_{xy}$ (in units of $a_0$), which conveniently approaches the value of zero at the BKT transition, along with a fit to the BKT scaling form (unfortunately, due to restrictions in the accessible system sizes, we were not able to explicitly follow this quantity to even lower temperatures than those shown in Fig.~\ref{fig:EPcoll}). The inset of Fig.~\ref{fig:EPcoll} provides the same data on a logarithmic scale, in order to more explicitly demonstrate the approach to a linear scaling of $\ln(a_0/\xi^\mathrm{EP}_{xy})$ with $1/\sqrt{T - T^\mathrm{EP}_\mathrm{BKT}}$, cf. Eq.~(\ref{eq:BKTxiscalinglog}), upon approaching $T^\mathrm{EP}_\mathrm{BKT}$. Based on our large-scale QMC simulations, we are thus able to access the correlation length scales that are required in order to observe the onset of  BKT scaling of the in-plane correlation length upon approaching the BKT transition of $H_\mathrm{EP}$, even for such a weak value of $D_z=0.01J$ as relevant for the compound BaNi${}_2$V${}_2$O${}_8$~\cite{Klyushina2017,Klyushina2021}. 

In the next section, we will examine to what extent this accessibility of the characteristic correlation length scaling near a BKT transition is affected by the additional presence of a finite in-plane easy-axis anisotropy $D_x>0$ in the full Hamiltonian $H$.

\section{Hierarchical Anisotropies}\label{Sec:HA}

After having established the BKT transition in the easy-plane limit, as well as the associated correlation length scaling,  we now consider the hierarchical model $H$ with finite values of both anisotropies. Since a finite value of $D_x>0$ breaks the O(2) symmetry from the easy-plane limit $H_\mathrm{EP}$ down to a residual $Z_2$ symmetry in the spin-X direction, the  hierarchical model $H$  exhibits a low-$T$ thermal Ising transition to a low-temperature antiferromagnetically ordered state, with a finite value of the staggered magnetization in the spin-X direction. In the following, we first focus
on the case of $D_x=0.1 D_z$, where $D_z=0.01J$ as in the previous section.
For this value of $D_x$, we are able  to examine the thermal phase transition in more detail than for even lower values of $D_x$ (due to the increasingly larger system sizes that are required to observe the asymptotic critical scaling at even lower values of $D_x$).
Later, we also turn to values of $D_x/D_z=0.01 - 0.05$, which are more relevant in view of
the compound BaNi${}_2$V${}_2$O${}_8$~\cite{Klyushina2017,Klyushina2021}.

\begin{figure}[t]
    \centering
    \includegraphics{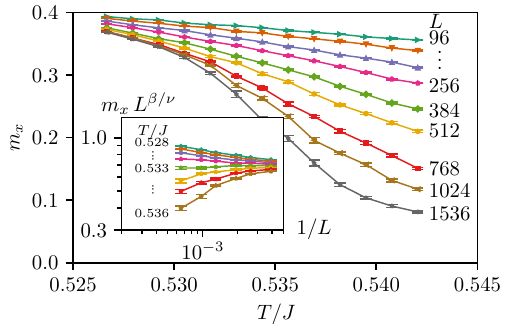}
    \caption{Order parameter estimator $m_x$ as a function of temperature $T$ for $D_z=0.01J$, $D_x=0.1D_z$ for different system sizes $L$ near the antiferromagnetic ordering transition. The inset shows a finite-size scaling plot to estimate $T_\mathrm{N}$,  based on the values  $\beta=1/8$ and $\nu=1$ for the 2D Ising universality class~\cite{Fisher1967}.}
    \label{fig:HA_mag2}
\end{figure}

In the QMC simulations,  we can quantify the emergence of the low-temperature magnetic state in terms of the  estimator
\begin{equation}
    m_x=\sqrt{\frac{1}{2L^2}S_x(\mathbf{0})}
\end{equation}
for the absolute value of the staggered magnetization. 
Fig.~\ref{fig:HA_mag2} shows the temperature dependence of $m_x$ for various linear system sizes $L$, for  $D_z=0.01 J$ and  $D_x=0.1 D_z$.
The temperature range in Fig.~\ref{fig:HA_mag2} focuses on the transition region into the low-$T$ ordered phase, which is seen to emerge in the thermodynamic limit below a N\'eel temperature of $T_\mathrm{N} \approx 0.53J$. As the antiferromagnetic order breaks the residual $Z_2$ symmetry of $H$, the thermal phase transition at $T_\mathrm{N}$ is expected to belong to the universality class of the 2D Ising model. Using the finite-size scaling $m_x\propto L^{-\beta/\nu}$ at criticality, with the exactly known values  $\beta=1/8$ and $\nu=1$ for the critical exponents in the 2D Ising universality class~\cite{Fisher1967} we extract
$T_\mathrm{N}/J=0.533(1)$, based on an appropriate scaling plot as shown in the inset of Fig.~\ref{fig:HA_mag2}.

\begin{figure}
    \centering
    \includegraphics{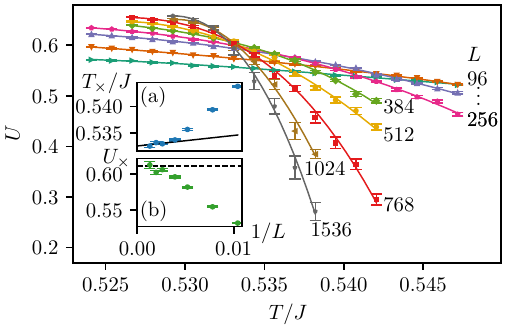}
    \caption{Binder cumulant $U$ as a function of temperature $T$ for $D_z=0.01J$, $D_x=0.1D_z$ for different system sizes $L$ near the antiferromagnetic ordering transition. 
    Close to the crossing points polynomials of degree 3 are used to interpolate the data (solid lines). Inset (a) shows the temperature $T_\times$ of the crossing points between the fitting polynomials of linear system sizes $L$ and $2L$ as functions of the inverse system size $1/L$, extrapolated to the thermodynamic limit using a linear fit (solid line). Inset (b) shows the value of the Binder cumulant $U_\times$ at the crossing points as a function of  $1/L$. The critical value $U_c$ of the Binder cumulant for an Ising transition on the triangular lattice is denoted by the dashed line.}
    \label{fig:HA_U}
\end{figure}

 In order to obtain a more accurate estimate for $T_\mathrm{N}$ and to confirm the anticipated Ising model universality class of the phase transition at $T_\mathrm{N}$, we analyze the Binder cumulant \cite{Binder1981,Binder1984}
\begin{equation}
    U = 1 - \frac{1}{3} \frac{\langle m_x^4\rangle}{\langle m_x^2 \rangle^2},
\end{equation}
which is shown in Fig.~\ref{fig:HA_U} for different system  sizes in the vicinity of $T_\mathrm{N}$. Using 
a crossing-point analysis of the values of $U$ for system sizes $L$ and $2L$, we can obtain an estimate 
for the ordering temperature: An  extrapolation of  the temperatures $T_\times$ of these crossing points to  the thermodynamic limit (cf. inset (a) of Fig.~\ref{fig:HA_U}), gives $T_\mathrm{N}/J=0.5325(1)$.  
The critical value of the Binder cumulant for the Ising model on triangular lattices with rhombic shapes has previously been determined to $U_c = 0.61182\dots$~\cite{Selke2005,Kamieniarz1993}. Given the underlying triangular structure of the honeycomb lattice, we expect the critical Binder ratio to agree with this value. We indeed find the data for the Binder cumulants at the crossing points, $U_\times$, to approach this value in the thermodynamic limit (cf. inset (b) of  Fig.~\ref{fig:HA_U}). This 
is in accord with  the expected universality class of the phase transition at $T_\mathrm{N}$. We note that for a controlled extrapolation to the thermodynamic limit we require rather large system sizes, e.g., the crossing points in the Binder ratio in Fig.~\ref{fig:HA_U} exhibit  significant drifts even for values of $L$ of several hundreds. 
Most importantly, we can confirm from this analysis that the phase transition at $T_\mathrm{N}$ belongs to the Ising universality class and thus the BKT transition, which takes place for $D_x=0$, is replaced by a true ordering transition in the full hierarchical model at finite $D_x$.

\begin{figure}[t]
    \centering
    \includegraphics{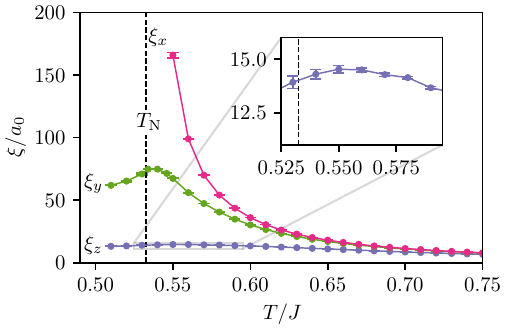}
    \caption{Correlation lengths $\xi_x$, $\xi_y$ and $\xi_z$ as functions of $T$ for $D_z=0.01J$, $D_x=0.1 D_z$. The dashed line indicates the N\'eel temperature $T_\mathrm{N}$.}
    \label{fig:HA_xis}
\end{figure}

After having established the thermal phase diagram of the hierarchical model, we now turn to analyze the behavior of the correlation lengths in this system. For this purpose, we determined the correlation lengths $\xi_x$, $\xi_y$, and $\xi_z$, in all three spin directions, using the approach from Sec.~\ref{Sec:XY}. Their temperature dependence, after an extrapolation to the thermodynamic limit, is shown in Fig.~\ref{fig:HA_xis}. While for large temperatures the three correlation lengths are very similar, as expected from the O(3) symmetry of the leading Heisenberg exchange term in $H$, 
they
exhibit noticeable different behavior below about 
 $T\approx 0.75J$. In particular,  both in-plane correlation lengths $\xi_x$ and $\xi_y$
 exhibit an enhanced further increase, whereas this is less pronounced for $\xi_z$. Similarly to the easy-plane case in Sec.~\ref{Sec:EP}, $\xi_z$  instead exhibits only a rather broad maximum at about $T\approx 0.55J$, i.e., slightly above $T_\mathrm{N}$, before its decrease in the ordered phase. We observe that $\xi_y$ still follows the increase of $\xi_x$ down to $T\approx 0.65J$.
 Within the temperature window 
 $0.65 \lesssim T/J  \lesssim 0.75$, the correlations can thus be characterized as easy-plane-like (note that this does not imply BKT scaling within this temperature regime). For even lower temperatures, $\xi_y$ however falls noticeably below $\xi_x$, and it reaches a maximum  at a similar temperature scale as $\xi_z$, but with a substantially larger maximum value, before it also decreases in the ordered phase in a more noticeable manner. Since $\xi_x$ is  the only diverging correlation length in the hierarchical system, we concentrate in the following on the behavior of this dominant correlation length  upon approaching the thermal phase transition.

\begin{figure}[t]
    \centering
    \includegraphics{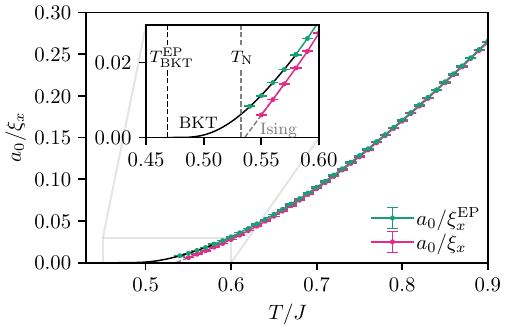}
    \caption{Inverse correlation length  $a_0/\xi_x$ for  $D_z=0.01J$, $D_x=0.1D_z$
    as a function of  temperature $T$, and compared to the inverse correlation length $a_0/\xi^\mathrm{EP}_x$ of the easy-plane model ($D_x=0$).  
    The solid line is a fit
    to the  exponential BKT scaling form 
    for $\xi^\mathrm{EP}_x$, and
    the dashed line  is an extrapolation of the linear drop in $a_0/\xi_x$ near the Ising transition.
    }
    \label{fig:HA_xi_fractions}
\end{figure}

While a finite value of the anisotropy $D_x>0$ strongly affects the nature of the phase transition and the scaling of  the correlation length $\xi_x$ in the low-temperature region, we expect it to be only little affected by the weak value of $D_x$ for temperatures well above $T_\mathrm{N}$. Indeed, we find  $\xi_x$ to closely follow  $\xi^\mathrm{EP}_{xy}$ at large temperatures. Upon approaching $T_\mathrm{N}$ however,  $\xi_x$ deviates increasingly from the easy-plane values, $\xi^\mathrm{EP}_{xy}$, as shown in
Fig.~\ref{fig:HA_xi_fractions}, where we again consider the inverse correlation lengths, since they conveniently approach zero at the thermal phase transitions. 
 
Upon lowering the  temperature we observe a different behavior in the transverse correlation lengths for the two models: while the data for the easy-plane model approaches the BKT scaling form (indicated by the solid line in Fig.~\ref{fig:HA_xi_fractions}),
the hierarchical model shows clear deviations from this behavior. Instead, we can identify (cf. the inset) the onset of a linear decrease of  $a_0/\xi_x$, which results from  the emergence of the algebraic scaling near the N\'eel temperature, $\xi_x\propto (T-T_\mathrm{N})^{-\nu}$ of the 2D Ising model universality, i.e., $\nu=1$. Indeed, the linear extrapolation of the linear drop in $a_0/\xi_x$, shown in the inset of Fig.~\ref{fig:HA_xi_fractions}, yields an upper bound for $T_\mathrm{N}$ that is only slightly larger than  the previously determined value of $T_\mathrm{N}$, where $a_0/\xi_x$ vanishes. 

\begin{figure}[t]
    \centering
    \includegraphics{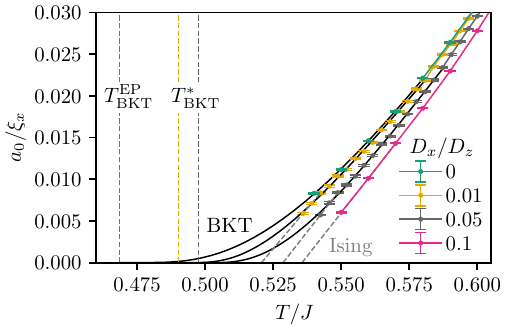}
    \caption{Inverse correlation length  $a_0/\xi_x$ for $D_z=0.01J$ and different values of $D_x$
    as functions of  temperature $T$, and compared to the inverse correlation length $a_0/\xi^\mathrm{EP}_x$ of the easy-plane model with $D_x=0$.  
    Solid black lines are fits
    to the  exponential BKT scaling form, and  dashed lines  extrapolations of the linear drop in $a_0/\xi_x$ near the Ising transition. 
    }
    \label{fig:HA_Dx_xi_fractions}
\end{figure}

We thus find that for the value of $D_x=0.1 D_z$ considered so far, the system does not show an extended crossover region separating the anisotropic high-$T$ region from the low-$T$ algebraic scaling of $\xi_x$ due to  the onset of the Ising criticality.
This situation is expected to change for even  smaller  value of $D_x$, since this weakens the effects of the in-plane anisotropy. More quantitatively, 
in Fig.~\ref{fig:HA_Dx_xi_fractions}, we compare the behavior of the correlation length  $\xi_x$ 
for varying values of $D_x$, as obtained from QMC simulations. We indeed find  that (i) for the lower two values of $D_x$, the data follows more closely the behavior of the easy-plane model towards lower temperatures, and (ii) for these lower values of $D_x$, we can identify an intermediate temperature regime in which the correlation length growth for the hierarchical model follows the BKT scaling prior to the onset of the asymptotic Ising scaling. More quantitatively, one can introduce an effective BKT transition temperature $T_\mathrm{BKT}^*$ (denoted $T_\mathrm{BKT}$ in Ref.~\cite{Klyushina2021}), such that the intermediate growth of $\xi_x$ can be fitted to the scaling in Eq.~(\ref{eq:BKTxiscaling})(with $T_\mathrm{BKT}^\mathrm{XY}$ replaced by $T_\mathrm{BKT}^*$), prior to the onset of the extrapolated characteristic linear Ising-model scaling of $a_0/\xi_x$ near the ordering transition. For both values of $D_x$,  
the extracted values of  $T_\mathrm{BKT}^*$ are smaller than the estimated values of $T_\mathrm{N}$, in accord with the interpretation that the BKT transition in  the easy-plane limit is  preempted by the onset of  N\'eel order, due to the finite value of $D_x$ in the full Hamiltonian $H$. Accordingly, upon lowering $D_x$, the value of 
$T_\mathrm{BKT}^*$ also approaches closer to the true BKT transition temperature $T_\mathrm{BKT}^\mathrm{EP}$ of the easy-plane limit.

Finally, we consider the estimation of the   N\'eel temperature for the lower values of $D_x$. As we already mentioned, it is not feasible to accurately determine $T_\mathrm{N}$ for these lower values of $D_x$ based on the analysis of the Binder cumulant that we performed for $D_x/D_z=0.1$, since inaccessibly large system sizes would be required for such an approach. 
One could then try to estimate 
$T_\mathrm{N}$ from the linear extrapolations shown in  Fig.~\ref{fig:HA_Dx_xi_fractions}. However, these extrapolations provide only an upper bound on $T_\mathrm{N}$, similar to what we observed already for the case of $D_x/D_z=0.1$ in Fig~\ref{fig:HA_xi_fractions}.
In fact, we expect that even lower temperatures (and therefore also larger system sizes -- due to the further increasing correlation length) are necessary in order to reach the asymptotic Ising scaling regime for $\xi_x$ and to reliably extract $T_\mathrm{N}$ from the extrapolation of the correlation length data for these low values of $D_x$.
In view of this limitation, it  would certainly  be interesting to quantify the actual $D_x$ dependence of $T_\mathrm{N}$ based on other, analytical treatments such as renormalization group calculations.

\section{Conclusions}\label{Sec:Conclusions}

We examined the thermal properties of anisotropic spin-1 Heisenberg antiferromagnets on the honeycomb lattice, with a focus on the behavior of the correlation length near the thermal phase transition. For this purpose, we first considered both the basic XY-model and the pure single-ion anisotropic easy-plane model. For both systems, we determined the value of the BKT transition temperature and also explored its $D_z$-dependence for the  easy-plane case. Furthermore, we confirmed that the correlation-length growth in the easy-plane case approaches the BKT scaling form upon approaching the BKT transition temperature. For the XY model, the BKT scaling is even observed up to temperatures at which the correlation length becomes of the order of the lattice constant. 

In addition, we considered the effects of a weak additional  in-plane easy-axis anisotropy, which breaks the O(2) symmetry of the pure easy-plane model down to a residual $Z_2$ symmetry. This provides us with a basic quantum spin model for examining the situation in the Ni${}^{2+}$ based  compound BaNi${}_2$V${}_2$O${}_8$. 
We were able to explicitly demonstrate the onset of Ising criticality for such a hierarchical model with two different single-ion anisotropies. However,  we also found that for sufficiently weak values of the easy-axis anisotropy,  as reported for BaNi${}_2$V${}_2$O${}_8$\cite{Klyushina2017},
one can still identify a narrow temperature regime above the N\'eel ordering temperature, in which the critical 
correlation length follows the characteristic BKT scaling form in terms of an effective BKT transition temperature $T_\mathrm{BKT}^*$, which lies between the N\'eel ordering temperature and the BKT transition temperature of the easy-plane limit. 

Returning to  the case of BaNi${}_2$V${}_2$O${}_8$, for which an extended BKT scaling regime was reported recently in the correlation length~\cite{Klyushina2021}, our results confirm that the characteristic BKT scaling  of the correlation length  can  be identified in the  hierarchical model 
of the magnetism in this compound
on length scales of the order of a  hundred lattice constants. It would of course  be important to more accurately quantify the width of this intermediate BKT scaling regime in terms of the hierarchical anisotropies. In addition, it would be interesting to take the  discrete lattice symmetries of BaNi${}_2$V${}_2$O${}_8$ into account in the microscopic modeling~\cite{Klyushina2017},  replacing thereby the residual $Z_2$ symmetry of $H$ by an $Z_6$ symmetry~\cite{Klyushina2021}, which is expected to further stabilize the BKT transition and its corresponding scaling regime~\cite{Jose1977}.

\section*{Acknowledgements}
We acknowledge insightful discussions with Bella Lake,  Ekaterina Klyushina and Johannes Reuther, as well as support by the Deutsche Forschungsgemeinschaft (DFG) through Grant No. WE/3649/4-2 of the FOR 1807 and through RTG 1995, and thank the IT Center at RWTH Aachen University and the JSC Jülich for access to computing time through the JARA Center for Simulation and Data Science.

\appendix

\section{Stochastic series expansion}\label{Sec:SSE}
The stochastic series expansion QMC method with directed loop updates \cite{Sandvik1991,Sandvik1999,Syljuasen2002,Alet2005} offers an unbiased approach to study sign-free quantum spin systems. 
In the following we  comment on some  technical aspects that are relevant for the SSE simulations of the  specific models that we considered here. For a more general and  detailed introduction, cf., e.g., Ref.~\cite{Syljuasen2002}.

The starting point of the SSE QMC method is a high temperature expansion of the partition function
\begin{equation}
    Z = \Tr \bigl( e^{-\beta H}\bigr) = \sum \limits_{\alpha} \sum \limits_{n=0}^\infty \frac{\beta^n}{n!} \bra{\alpha}(-H)^n\ket{\alpha},
    \label{eq:partfun}
\end{equation}
where $\{\ket{\alpha}\}$ is a orthonormal basis of the Hilbert space of $H$, called the computational basis. Here, we use the standard local product $S^z$ basis, i.e.,  $\ket{\alpha} = \ket{S_1^z,S_2^z,\dots,S_N^z}$. To evaluate the matrix elements $\bra{\alpha} (-H)^n \ket{\alpha}$, the Hamiltonian $H$ is decomposed as  $H = -\sum_{b,t} H_{b,t}$ into a sum of bond operators $H_{b,t}$, specified by a bond index $b$, and the operator type  $t$. These bond operators must be non-branching, i.e.,  the action of $H_{b,t}$ on a given basis state $\ket{\alpha}$ is proportional to another  basis state $\ket{\alpha^\prime}$. Introducing a sequence of bond operators $S_n = \{[b_1,t_1],\dots,[b_n,t_n]\}$ that contributes to the partition function, we can rewrite $Z$ as
\begin{equation}
    Z = \sum \limits_{\alpha} \sum \limits_{n=0}^\infty \sum_{\{S_n\}} \frac{\beta^n}{n!} \bra{\alpha}\prod \limits_{i=1}^n (H_{b_i,t_i})\ket{\alpha},
    \label{eq:partfun_bond}
\end{equation}
where the expansion order $n$ corresponds to the number of operators in $S_n$, i.e., its length. 
In practice, the expansion order is fixed to some cut-off $L$ that is set larger than the maximally sampled expansion order. In this fixed length representation the operator string $S_L$ is padded with unity operators, such that $n$ corresponds to the number of non-unity operators in $S_L$. The expansion order $n$, the state $\ket{\alpha}$ as well as the operators in the string $S_L$ are then sampled during the Monte Carlo updating procedures.

In the diagonal update step, operators  that are diagonal in the computational basis are inserted or removed from $S_L$. The second update step is the directed-loop update, which is a global update that proceeds via locally constructing a cluster of operators in $S_L$, viewed as list of vertices along with two incoming and outgoing legs. The latter carry the local spin state of the two sites that belong to the bond $b$ of the bond operator. If the matrix elements of the diagonal operators are much larger than those of the offdiagonal operators, these local steps during the (global) directed-loop update each have very low acceptance probabilities. This can cause the updating dynamics to freeze and may lead to ergodicity problems of the Monte Carlo update.

After these general remarks, we consider the  hierarchical Hamiltonian $H$, i.e., 
\begin{equation}
    H = J \sum \limits_{\langle i,j \rangle} \mathbf{S}_i \cdot \mathbf{S}_j + D_z \sum \limits_i (S_i^z)^2 - D_x \sum \limits_i (S_i^x)^2 ,
\end{equation}
and we first examine the decomposition into bond operators. In the following, we consider a bond $b$ connecting the two sites $i$ and $j$ on the honeycomb lattice. The  decomposition leads to  three types of bond operators $H_{b,t}$, $t=0,1,2$, which  are classified by the action on the spins connected by this bond: (i) a bond operator that is diagonal in the computational  basis, 
\begin{equation}
    H_{b,0} = C -J S_i^z S_j^z - \sum \limits_{k \in \{ i,j \} } \biggl(\frac{D_z}{z}(S_k^z )^2 - \frac{D_x}{4 z} (S_k^+ S_k^- + S_k^- S_k^+)\biggr),
\end{equation}
where an appropriate constant $C$ can be added in order to ensure that all matrix elements are positive, 
(ii) a first offdiagonal part due to the Heisenberg exchange,
\begin{equation}
    H_{b,1} =- \frac{J}{2}(S_i^+S_j^- + S_i^-S_j^+),
\end{equation}
as well as (iii) a second offdiagonal part due to the easy-axis  anisotropy $D_x$,
\begin{equation}
    H_{b,2} = \frac{D_x}{4 z} \sum \limits_{k \in \{ i,j \}} (S_k^+S_k^+ + S_k^-S_k^-).
\end{equation}
Here, $z=3$ is the coordination number on the honeycomb lattice. The full Hamiltonian is given by the  sum  $H =N_b C -\sum_{b,t} H_{b,t}$ over these bond operators ($N_b$ denotes the number of bonds on the finite lattice). On a bipartite lattice, such as the honeycomb lattice considered here, all finite contributions to the partition function in Eq.~(\ref{eq:partfun_bond})  have positive weights, and can thus be sampled without a sign problem. 

Several observations are in order: 
(i) For finite values of $D_x$, the presence of the bond operator $H_{b,2}$ leads to the following modification from   the standard directed loop update: the head of the moving operator, which is assigned a local $S^+$ or $S^-$ operator, is now allowed to switch-and-revert~\cite{Syljuasen2002} to the other site of a local vertex without being inverted. 
(ii)
The easy-axis anisotropy $D_x$ contributes to both the diagonal and offdiagonal operators, whereas the easy-plane anisotropy $D_z$ contributes only to diagonal operators. In the limit of large $|D_z|$,  this leads to sampling problems, which result in larger statistical errors. We observed that these are reduced, if the larger of the two anisotropies aligns in the spin-X direction. For this purpose, one can perform a rotation of the Hamiltonian in the spin plane about the spin-Y axis, without introducing a QMC sign problem due to the bipartiteness of of the honeycomb lattice. (iii) 
Some observables, in particular the Binder ratio $U$, are more readily accessible after performing such a rotation of the Hamiltonian about the spin-Y axis. Indeed,  the second and forth moments of the the order parameter are then  diagonal observables in the computational basis.
For our simulations, we took those observations into account in order to optimize the computational efforts. 

\section{Quantum phase transition in the pure easy-plane model}\label{Sec:EP_app}
In this appendix, we examine the high-$D_z$ continuous quantum phase transition of the spin-1 easy-plane Hamiltonian 
\begin{equation}
    H_\mathrm{EP} = J \sum \limits_{\langle i,j \rangle} \mathbf{S}_i \cdot \mathbf{S}_j + D_z \sum \limits_i (S_i^z)^2 
\end{equation}
in more detail.
In addition to breaking the SO(3) symmetry of the Heisenberg model, finite values of $D_z > 0$ suppress the local spin states $S_i^z = \pm 1$, whereas the local state $S_i^z= 0$ is preferred by finite $D_z>0$. In the large-$D_z$ limit, the ground state is given by the direct product state $\ket{0}=\prod_i\ket{S_i^z=0}$, with $S_i^z=0$ on each lattice site $i$. We expect a quantum phase transition to take place at a finite value of $D_z>0$, beyond which the XY-antiferromagnetic ground state gets replaced by a non-magnetic state that connects to this large-$D_z$ limit product state. To gain insight into this transition, we can employ a simple perturbative argument in the limit $J/D_z \ll 1$: The energy spectrum of the unperturbed Hamiltonian $H^{(0)} = D_z \sum \limits_i (S^z_i)^2$ in this limit is described by the number of local  $S_i^z = \pm 1$ states, which we denote by $N_\pm$. It is therefore given by the discrete energies $E^{(0)}_{N_\pm}=D_z N_\pm$, which are  well separated from each other. 
The direct product state $\ket{0}$ is the ground state of $H^{(0)}$, with $E^{(0)}_0=0$. The lowest excited states belong to the  $N_\pm = 1$ sector with $E^{(0)}_1=D_z$. In contrast to the ground state $\ket{0}$, this energy level is thus highly degenerate. The Heisenberg exchange interaction allows a local $S_i^z = \pm 1$ excitation atop the state $|0\rangle$ to hop on the honeycomb lattice. This leads to an effective tight-binding kinetic energy contribution, with a hopping amplitude that is equal to $J$ on the honeycomb lattice.   
This results into a $J$-dependent change of the lowest excitation energy in the 
$N_{\pm}= 1$ sector to $E^{(0)}_1+E^{(1)}_1=D_z-3J$ on the three-fold coordinated honeycomb lattice within this first-order perturbation theory.  
\begin{figure}[t]
    \includegraphics{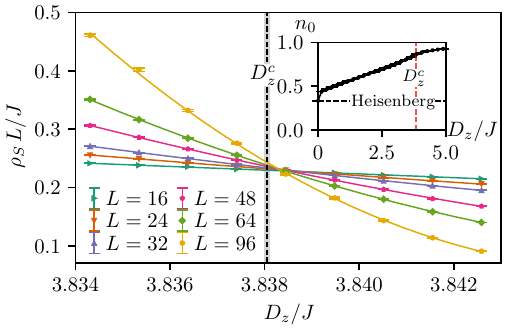}
    \caption{Spin stiffness $\rho_S$ multiplied by $L$ at $\beta=2L$ as a function of $D_z$ of the spin-1 easy-plane model $H_\mathrm{EP}$  on the honeycomb lattice of different linear system size $L$. 
    Polynomials of degree 3 are used to interpolate the data (solid lines).
    The critical value $D_z^\mathrm{c}$, obtained by extrapolating the crossing points of the fitting polynomials in $\rho_S L$ for system sizes $L$ and $2L$, is denoted by the (dashed) vertical line. The inset shows the mean occupation density $n_0$ of the local $S_i^z = 0$ states as a function of $D_z$ (for $L=48$, $\beta=2L$). The  vertical line denotes $D_z^\mathrm{c}$, while the  horizontal lines indicates the occupation density of the spin-1 Heisenberg model.}
    \label{fig:EP_QCP}
\end{figure}

The direct product state $\ket{0}$ in the $N_\pm=0$ sector instead does not change within first-order perturbation theory.
Upon comparing the energies from the two sectors, we thus expect from this lowest-order perturbative calculation  a critical value $D_z^c$ of $D_z$, where a transition takes place out of the large-$D_z$ ground state $\ket{0}$.
More specifically, we obtain from this analysis a first-order estimate for the critical easy-plane anisotropy of $D_z^c=3J$. A more quantitative computation needs to consider also higher-order terms and excited states in the perturbative expansion, which result in a dressing of the ground state $\ket{0}$ for finite values of $J/D_z$. We furthermore expect those contributions to replace the 
level-crossing transition from the first-order approach by a continuous quantum phase transition in the thermodynamic limit. Indeed, from symmetry considerations, we expect a continuous quantum phase transition to separate the two regimes, belonging to the three-dimensional O(2) universality class, based on the  global O(2) symmetry of the easy-plane Hamiltonian in $d=2$ spatial dimensions. In addition, the dynamical critical exponent is then equal to $z=1$.
In order to accurately locate the quantum phase transition, we turn to QMC simulations. More specifically, we consider the spin stiffness $\rho_S$, which scales at the  quantum critical point as ~\cite{Josephson1966,Fisher1973,Troyer1997,Sandvik2010}
\begin{equation}
    \rho_S \propto L^{2-d-z},
\end{equation}
in order to accurately calculate $D_z^c$. 
Based on $z=1$, we measured $\rho_S$ at an inverse temperature of $\beta=2L$ for different linear system sizes up to $L=96$ and for different easy-plane anisotropies $D_z$. The numerical results for $\rho_S$  are shown in Fig.~\ref{fig:EP_QCP}. The value of $D_z^c/J= 3.83805(5)$ is then obtained upon extrapolating  crossing points in $\rho_S$ from system sizes $L$ and $2L$ to the thermodynamic limit, as indicated by the vertical line in Fig.~\ref{fig:EP_QCP}.

The actual value of $D_z^c$ is larger that the above first-order perturbative estimate, which however already provides  the right order of magnitude.
Also included in Fig.~\ref{fig:EP_QCP} (in the inset) are  QMC results for the mean occupation  density $n_0$
of the local $S_i^z = 0$ states. This quantity increases from the exact SU(2)-symmetric value of $1/3$ for $D_z=0$ to the limiting value of $1$ in the large-$D_z$ limit. Furthermore, it evolves smoothly across the quantum phase transition (as befits a continuous transition), merely exhibiting a mild kink at the quantum critical point. 

Near the quantum critical point at $D_z^c$, the BKT transition temperature is expected to scale as 
\begin{equation}\label{Eq:nuzscal}
 T_\mathrm{BKT}^\mathrm{EP}\propto (D-D_z^c)^{\nu z}, 
\end{equation}
where $\nu=0.67155(27)$ for the 3D XY model~\cite{Campostrini2001}.
\begin{figure}[t]
    \includegraphics{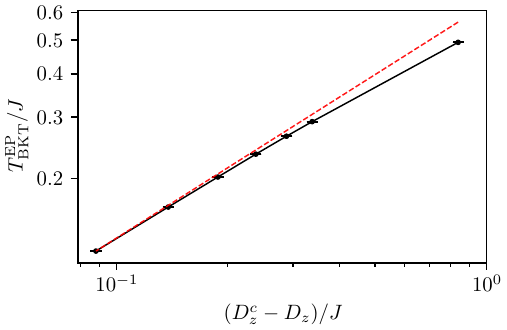}
    \caption{BKT transition temperature of the spin-1 easy-plane model $H_\mathrm{EP}$ on the honeycomb lattice in the vicinity of the quantum critical point in a log-log plot, compared to the scaling prediction in Eq.~(\ref{Eq:nuzscal}) near $D^c_z$, indicated by the slope of the dashed line.}
    \label{fig:EP_Tscal}
\end{figure}
As shown in Fig.~\ref{fig:EP_Tscal}, our numerical results for $T_\mathrm{BKT}^\mathrm{EP}$
are in accord  with an approach to this scaling close to  $D_z^c$. 

\section{Pure easy-axis model}\label{Sec:EA}
In this appendix, we examine the pure easy-axis regime ($D_z=0$) of the Hamiltonian $H$. Moreover, we also introduce an exchange anisotropy in the form of
the additional parameter $\lambda$, such that we consider here the  Hamiltonian
\begin{equation}
    H_\mathrm{EA}=J\sum_{\langle i,j \rangle} S_i^x S_j^x + \lambda  (S_i^y S_j^y + S_i^z S_j^z) - D_x \sum_i (S^x_i)^2, 
\end{equation}
which for $\lambda=1$ recovers the original  Heisenberg interaction of $H$, while in the limit $\lambda=0$ a classical spin model is obtained. This classical limit of our spin-1 model is the well known Blume-Capel model~\cite{Blume1966,Capel1966}. Both models are usually formulated in terms of the spin-Z direction as the easy axis  instead of the spin-X direction, which we use here in order to remain consistent with our convention in the main part of the paper. 

\begin{figure}[t]
    \centering
    \includegraphics{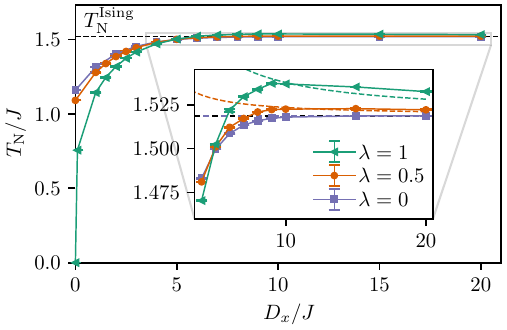}
    \caption{Néel temperature $T_\mathrm{N}$ as a function of the easy-axis anisotropy $D_x$ in $H_\mathrm{EA}$ for different values of $\lambda$. The dashed lines denote  results  from second-order Brillouin-Wigner perturbation theory in the large $D_x$-limit.}
    \label{fig:EA_results}
\end{figure}

As in the full hierarchical model $H$, a finite value of $D_x>0$ leads to a thermal phase transition into a low-$T$ phase with antiferromagnetic order in the spin-X direction. We can obtain the corresponding N\'eel temperature $T_\mathrm{N}$ from QMC simulations based on the Binder cumulant analysis as discussed in Sec.~\ref{Sec:HA}  (for $\lambda=0$, we used the approach of Ref.~\cite{Zierenberg2017} to  simulate the Blume-Capel model). The results for $T_\mathrm{N}$ for different values of $\lambda$ are summarized in Fig.~\ref{fig:EA_results}.

Let us first consider the classical (Blume-Capel) limit, $\lambda=0$. Here, for large values of $D_x$, the N\'eel temperature approaches the value of the two-dimensional Ising model (on the honeycomb lattice), which is  known exactly and equal to $T_\mathrm{N}^\mathrm{Ising}/J=1.518\dots$ \cite{Houtappel1950}.
Indeed, a large value of $D_x$ suppresses the local spin states $S_i^x=0$ (in the spin-X basis), whereas the local spin states $S_i^x =\pm 1$ are energetically  favorable. As a result, for $\lambda=0$, due to the $S_i^x S_j^x$ term in the exchange coupling, we  (exactly) obtain an Ising model in the limit $D_x \to \infty$. 

Turning now to the case of non-zero $\lambda>0$, we find that again the N\'eel temperature tends towards the  Ising model value for large $D_x$. However, we observe in this case a non-monotonous behavior in the $D_x$-dependence of $T_\mathrm{N}$, as seen in the zoom of Fig.~\ref{fig:EA_results}. In particular, we find that for finite values of $\lambda$, the N\'eel temperature approaches 
$T_\mathrm{N}^\mathrm{Ising}$ from above, in contrast to the classical limit $(\lambda=0)$, in which 
$T_\mathrm{N}$ increases monotonously and approaches
$T_\mathrm{N}^\mathrm{Ising}$  
from below.

To gain analytical insight into this behavior, we investigated the  large-$D_x$ region using second-order Brillouin-Wigner perturbation theory to derive an effective Hamiltonian in the large-$D_x$ regime. We find that, up to second order in $J/D_x$, $H_\mathrm{EA}$ can be described by a spin-1/2 XXZ model, given by
\begin{equation}
    H_\mathrm{EA}^{\mathrm{eff}} = J \sum \limits_{\left<i,j\right>}
    -\lambda_\mathrm{eff}(S_i^yS_j^y + S_i^zS_j^z) + \Delta S_i^xS_j^x,
\end{equation}
where
\begin{equation}\label{Eq:AD}
    \lambda_\mathrm{eff} = \frac{\lambda^2}{2}\frac{J}{D_x}, \qquad \Delta = \biggl(4  + \frac{\lambda^2}{2}\frac{J}{D_x}\biggr).
\end{equation}
The derivation of this effective Hamiltonian can be found in Appendix~\ref{Sec:EA_pert}. 

In terms of the parameters of 
$H_\mathrm{EA}^{\mathrm{eff}}$, the  large-$D_x$ regime corresponds to the  limit $\Delta \gg \lambda_\mathrm{eff}$, in which  
the N\'eel temperature of the spin-1/2 XXZ model approaches 
$\Delta T_\mathrm{N}^{\mathrm{Ising}} /4$ (cf. also Ref.~\cite{Goettel2012} for a square lattice geometry). From the expression for $\Delta$ in Eq.~(\ref{Eq:AD}), we thus indeed find that for finite $\lambda>0$ in the large-$D_x$ regime,
the N\'eel temperature
\begin{equation}
T_\mathrm{N}\approx \biggl(1  + \frac{\lambda^2}{8}\frac{J}{D_x}\biggr)T_\mathrm{N}^{\mathrm{Ising}}
\end{equation}
approaches $T_\mathrm{N}^\mathrm{Ising}$ from above, as observed also in the QMC data. 
A quantitative comparison between the QMC data and the perturbation theory result is included in
Fig.~\ref{fig:EA_results}. We find that the second-order perturbation theory fits well to the trend seen in the QMC data for  increasingly large values of $D_x$. However, this comparison also reveals  that higher order contributions become important at lower values of $D_x$, as the crossover to decreasing N\'eel temperatures at smaller $D_x$ cannot be captured in this order of perturbation theory.

\section{Brillouin-Wigner perturbation theory}\label{Sec:EA_pert}

Here, we detail the Brillouin-Wigner perturbation theory \cite{Brillouin1932,Wigner1935,Messiah1962} that we used to derive an effective Hamiltonian in the Ising limit $J\ll D_x$ of the easy-axis Hamiltonian $H_\mathrm{EA}$. In this appendix, we work in the $S^z$ basis, and therefore we first rotate the Hamiltonian such that the easy-axis anisotropy aligns in the spin-Z direction.  Expressing the rotated Hamiltonian in units of the easy-axis anisotropy $D_x$, we obtain
\begin{equation}
    \tilde{H} =\underbrack{\frac{J}{D_x} \sum \limits_{\langle i,j\rangle} \lambda (S_i^x S_j^x + S_i^y S_j^y) + S_i^z S_j^z}_{\text{Perturbation } V\text{ with }J/D_x\ll 1} - \underbrack{\sum \limits_{\phantom{\langle}i\phantom{\rangle}} S_i^z S_i^z}_{H^{(0)}}.
\end{equation}
 The energy spectrum of the unperturbed part $H^{(0)}$ is given  by the number of $\ket{0}$ states $N_0$ as $E^{(0)} = -(N - N_0)$. Therefore, the subspaces of the Hilbert space with different number of $\ket{0}$ states $N_0$ are well separated compared to $J/D_x\ll 1$. In the Ising limit $J/D_x \to 0$, the lowest energy subspace has $N_0=0$.  We thus  divide the Hilbert space into the subspace with $N_0=0$ (subspace 1) and $N_0\geq 1$ (subspace 2). The Schrödinger equation in this notation is given by
\begin{equation}
  \begin{pmatrix}
    H_{11}^{(0)} + V_{11} & V_{12} \\
    V_{21} & H_{22}^{(0)} + V_{22}
  \end{pmatrix}
  \begin{pmatrix}
    \psi_1\\
    \psi_2
  \end{pmatrix}
  =
  E
  \begin{pmatrix}
    \psi_1\\
    \psi_2
  \end{pmatrix} ,
\end{equation}
where $H_{ii} = H_{ii}^{(0)} + V_{ii}$ describe the Hamiltonians in  subspace $i$ and $V_{12}$ and $V_{21}$ are perturbations that couple the subspaces 1 and 2. This gives the two equations
\begin{align}
      H_{11} \psi_1 + V_{12}\psi_2 = E\, \psi_1 ,\\
  V_{21}\psi_1 + H_{22} \psi_2 = E\psi_2  .
\end{align}
To obtain an effective theory in subspace 1 we can insert the second equation in to the first and get
\begin{equation}
      H_{11}^{\mathrm{eff}} = H_{11}^{(0)} + V_{11} + V_{12} \frac{1}{E-H_{22}^{(0)} - V_{22}} V_{21} .
\end{equation}
The energy dependence of the 
 Hamiltonian $H_{11}^\mathrm{eff}$ can be eliminated by expanding the energy $E = \sum \limits_{k=0}^{\infty}E^{(k)}$ with $E^{(k)}\propto \mathcal{O}\bigl((J/D_x)^k\bigr)$. Using the identity
\begin{equation}
    \frac{1}{A-B} = \sum \limits_{k=0}^{\infty} \Bigl( \frac{1}{A} B\Bigr)^k \frac{1}{A},
\end{equation}
with $A=E^{(0)} - H_{22}^{(0)}$ and $B=V_{22} - \sum \limits_{k=1}^{\infty} E^{(k)}$ then yields the general form
\begin{multline}
    H_{11}^{\mathrm{eff}} = H_{11}^{(0)} + V_{11} +\\ \sum \limits_{l=0}^{\infty} V_{12} \biggl[ \frac{1}{E^{(0)}-H_{22}^{(0)}} \Bigl(V_{22} - \sum \limits_{k=1}^\infty E^{(k)} \Bigr)\biggr]^l\times \\ \times \frac{1}{E^{(0)}- H_{22}^{(0)}} V_{21}.
\end{multline}
Therefore, the effective Hamiltonian up to second order is given by
\begin{equation}
  H_{11}^{\mathrm{eff}} = H_{11}^{(0)} + V_{11}+
  V_{12} \frac{1}{E^{(0)}-H_{22}^{(0)}} V_{21} +
  \mathcal{O}\biggl(\Bigl(\frac{J}{D_x}\Bigr)^3\biggr) .
\end{equation}
 The first-order correction $V_{11}$ is given by
\begin{equation}
    V_{11} = \frac{J}{D_x} \sum \limits_{\langle i,j\rangle} P_1 (S_i^zS_j^z) P_1,
\end{equation}
where $P_1$ is a projector onto the subspace 1. This can be expressed using classical Ising spins $\sigma_i=\pm 1$, such that $V_{11}=(J/D_x)\sum \limits_{\langle i,j\rangle} \sigma_i\sigma_j$. The offdiagonal part of the perturbation $V_{12}$, that couples the subspaces 1 and 2, is given by
\begin{equation}
    V_{21} = \sum \limits_{\langle i,j \rangle} P_2 \frac{J}{D_x} \frac{\lambda}{2} (S_i^+S_j^- + S_i^-S_j^+) P_1, 
\end{equation}
and $V_{12}$ analogously with just the order of the projectors changed. This yields for the second-order correction
\begin{multline}
  V_{12} \frac{1}{E^{(0)} - H_{22}^{(0)}}V_{21} = \\ P_1 \sum
  \limits_{\left<m,n\right>} \sum
  \limits_{\left<i,j\right>}\Bigl(\frac{J}{D_x}\Bigr)^2\frac{\lambda^2}{4}\frac{1}{-N
  - (-\sum \limits_k S_k^zS_k^z)}\times \\ \times (S_m^+S_n^- + S_m^-S_n^+)(S_i^+S_j^- +
  S_i^-S_j^+)P_1  .
\end{multline}
Each virtual bond state that is created in subspace 2 has to be acted on again to get back to subspace 1, thus yielding $\langle m,n\rangle=\langle i,j \rangle$. For each bond state $\ket{\pm 1, \mp 1}$, the action of the operators creates the virtual state $\ket{0,0}$ and act on it again, so we obtain the processes
\begin{equation}
  \ket{\pm 1, \mp 1} \to \ket{0,0}
  \begin{casesnew}[-stealth]
    \ket{\pm 1, \mp 1} \\
    \ket{\mp 1, \pm 1}
  \end{casesnew}.
\end{equation}
If the bond states in subspace 1 are parallel $\ket{\pm 1, \pm 1}$ the action of the operators is 0. Therefore, the virtual state differs for each bond by exactly two $\ket{0}$ states, so that
\begin{equation}
    \frac{1}{E^{(0)} - H_{22}^{(0)}} = \frac{1}{-N - (-(N-2))} = -\frac{1}{2} .
\end{equation}
We thus obtain for the second order correction
\begin{equation}
    H_{11}^{(2)}=-\frac{\lambda^2}{8}\biggl(\frac{J}{D_x}\biggr)^2 P_1(V_\mathrm{diag} + V_\mathrm{offdiag})P_1,
\end{equation}
where we introduced the offdiagonal part 
\begin{equation}
    V_\mathrm{offdiag} = \sum \limits_{\langle i,j \rangle} (S_i^{+})^2(S_j^{-})^2 + (S_i^{-})^2(S_j^{+})^2,
\end{equation}
and the diagonal part 
\begin{equation}
    V_\mathrm{diag} =\sum \limits_{\langle i,j \rangle} S_i^-S_i^+S_j^+S_j^- +  S_i^+S_i^-S_j^-S_j^+ .
\end{equation}
We now consider $V_\mathrm{diag}$ and $V_\mathrm{offdiag}$ in more detail. Both operators can be expressed in terms of spin-1/2 degrees of freedom. First, we examine $V_\mathrm{diag}$, which can be expressed as
\begin{align}
    V_\mathrm{diag} &=  2 \sum \limits_{\langle i,j \rangle} \delta_{\sigma_i,\sigma_j} - \sigma_i \sigma_j = 2 \sum \limits_{\langle i,j \rangle} \frac{1}{2}(\sigma_i\sigma_j + 1) - \sigma_i \sigma_j \nonumber \\
    &= N_b - 4 \sum \limits_{\langle i,j \rangle} S_i^z S_j^z,
\end{align}
where $\sigma_i=\pm 1$ are as previously introduced classical Ising spins, $S_i^z$ are spin-1/2 variables, and $N_b$ is the number of bonds on the lattice. Turning our attention to the offdiagonal part, we can express it by spin-1/2 operators as follows 
\begin{equation}
    V_\mathrm{offdiag} = 2 \sum \limits_{\langle i,j \rangle} (S_i^+S_j^- + S_i^- S_j^+) .
\end{equation}
Previously we saw that the first order correction $V_{11}$ is a classical Ising model with coupling $J/D_x$. This can be expressed in terms of spin-1/2 variables as well, such that $V_{11} = (4J/D_x) \sum \limits_{\langle i,j \rangle} S_i^z S_j^z$.

Finally, taking into account the first and second order corrections and expressing the Hamiltonian in its original units we obtain the effective spin-1/2 Hamiltonian in the subspace 1
\begin{equation}
    H_{11}^{\mathrm{eff}} = J\sum \limits_{\left<i,j\right>}
    -\frac{\lambda_\mathrm{eff}}{2}(S_i^+S_j^- + S_i^-S_j^+) +
    \Delta S_i^zS_j^z,
\end{equation}
where 
\begin{equation}
    \lambda_\mathrm{eff} = \frac{\lambda^2}{2}\frac{J}{D_x}, \qquad \Delta = \biggl(4  + \frac{\lambda^2}{2}\frac{J}{D_x}\biggr).
\end{equation}
In the large $D_x$ limit, the easy-axis Hamiltonian can therefore be described by an effective spin-1/2 XXZ model, where in the limit $J/D_x \to 0$, irrespective of the value of $\lambda$, the Ising model is obtained exactly.

\bibliography{references.bib}

\end{document}